# Correlation between Dzyaloshinskii–Moriya interaction and spin mixing conductance at antiferromagnet / ferromagnet interface

Xin Ma[1,*], Guoqiang Yu[2], Seyed A. Razavi[3], Liang Chang[1], Lei Deng[1], Zhaodong Chu[4], Congli He[3], Kang L. Wang[3, 5], and Xiaoqin Li[4]

[1]*Department of Electrical and Computer Engineering, University of California, Santa Barbara, California 93106, USA*
[2]*Beijing National Laboratory for Condensed Matter Physics, Institute of Physics, Chinese Academy of Sciences, Beijing 100190, China*
[3]*Department of Electrical Engineering, University of California, Los Angeles, California 90095, USA*
[4]*Department of Physics, Center for Complex Quantum systems, The University of Texas at Austin, Austin, Texas 78712, USA*
[5]*Department of Physics, University of California, Los Angeles, California 90095, USA*

The rich interaction phenomena at antiferromagnet (AFM)/ ferromagnet (FM) interfaces are key ingredients in AFM spintronics, where many underlying mechanisms remain unclear. Here we report a correlation observed between interfacial Dzyaloshinskii–Moriya interaction (DMI) $D_S$ and effective spin mixing conductance $g_{eff}^{\uparrow\downarrow}$ at IrMn/CoFeB interface. Both $D_S$ and $g_{eff}^{\uparrow\downarrow}$ are quantitatively determined with Brillouin light scattering measurements, and increase with IrMn thickness in the range of 2.5~7.5 nm. Such correlation likely originates from the AFM-states-mediated spin-flip transitions in FM, which promote both interfacial DMI and spin pumping effect. Our findings provide deeper insight into the AFM-FM interfacial coupling for future spintronic design.

The antiferromagnet (AFM)/ ferromagnet (FM) interfaces are of central importance in the recent development of AFM spintronics [1-5]. Through the interfacial coupling, the unique electric, magnetic and transport properties of the AFM can be bridged to control the FM layer. For instance, the adjacent AFM layer improves the hardness of FM via exchange bias (EB) [6-8] or enhances the spin current transport away from FM [9-13]. Taking advantage of the faster dynamics in AFM, one can speed up the optical control of FM by selectively perturbing the spin arrangement of the neighboring AFM layer [5]. An intense and transient torque is subsequently generated onto the FM across the AFM/FM interface [5]. More recently, new strategies utilizing multiple interfacial interactions in synergy lead to promising technology breakthroughs. Examples include the pure electric switching of FM magnetization [1-4] and the establishment of magnetic skyrmions in AFM/FM systems [14]. Especially, the electric current induced magnetization switching is driven by the spin-orbit torque (SOT) generated in the AFM or at the AFM/FM interface [1, 2, 15-17], which also utilizes EB instead of the external magnetic field to break the switching symmetry [1-4]. In addition, magnetic skyrmion phase has been stabilized at room temperature in AFM/FM systems [14], resulting from the interplay with Dzyaloshinskii–Moriya interaction (DMI), interfacial magnetic anisotropy and EB. The directional motion of such Néel-type magnetic skyrmions can also be efficiently manipulated with the SOT in AFM/FM systems [14].

Among the rich interactions at AFM/FM interface, the recently observed interfacial DMI remains most puzzling. While such DMI at AFM/FM interface also promotes non-collinear spin alignments, it exhibits important difference from that in heavy metal (HM)/FM bilayers investigated extensively in recent years [18-22]. Notably, the DMI at IrMn/CoFeB interface can be enhanced by increasing the IrMn thickness well beyond the spin diffusion length [23], overcoming a bottleneck for improving DMI via increasing the HM layer thickness in the HM/FM bilayers [24, 25]. In light of DMI's important role in varied spintronic applications [26-29], elucidating the DMI across the AFM-FM interface is not only important from a scientific point of view, but also of great technologic relevance.

In this Letter, we aim to provide deeper insights into the newly observed DMI at AFM/FM (IrMn/CoFeB) interface, especially such DMI's intriguing dependence on the IrMn thickness $t_{\text{IrMn}}$ [23]. We characterized the effective spin mixing conductance $g_{\text{eff}}^{\uparrow\downarrow}$ at IrMn/CoFeB interfaces from the magnetic field dependence of linewidth broadening in Brillouin light scattering (BLS) measurements. Both interfacial DMI strength $|D_S|$ and $g_{\text{eff}}^{\uparrow\downarrow}$ continuously increase when $t_{\text{IrMn}}$ increases from 2.5 to 7.5 nm in the IrMn/CoFeB/MgO multilayer thin films. We use such correlation to elucidate the underlying physics of the DMI at IrMn/CoFeB interface, with the help of the better understood spin pumping effect. The surprising enhancement of DMI with larger $t_{\text{IrMn}}$ likely originates from the enlarged spin-orbit coupling (SOC) strength of Mn $3d$ states around the Fermi level and their facilitation on spin-flip transitions in the CoFeB layer, which is manifested by the increase of $g_{\text{eff}}^{\uparrow\downarrow}$. Our discovery is in synergy with many on-going activities investigating the correlation between DMI and other SO effects including SOT [30-32], proximity induced magnetization [25, 33, 34], and magnetic anisotropy [35].

The $\text{Ir}_{22}\text{Mn}_{78}(t)/\text{Co}_{20}\text{Fe}_{60}\text{B}_{20}(2)/\text{MgO}(2)/\text{Ta}(2)$ multilayer thin films were deposited by magnetron sputtering at room temperature on thermally oxidized silicon substrates, where the subscript represents the percentage of each element in the alloyed layer and the numbers in parentheses denote the collected and sent to a Sandercock-type multipass tandem Fabry-Perot interferometer. In order to guarantee a high-quality spectra lineshape and minimize the uncertainty in magnon momentum space, the BLS probe area is about 100 μm in diameter and an additional spatial filter was placed in the signal collection path. On one hand, DMI was quantitatively measured from momentum-resolved BLS experiment by varying the incident angle of light $\theta$, where such approach has been demonstrated by many groups [24, 39-43]. On the other hand, BLS measurements with a fixed incident angle $\theta = 45°$ were conducted to derive the spin pumping enhanced magnetic damping $\alpha_{sp}$. In such a probe geometry, the magnon-magnon scatterings' contribution to linewidth broadening of BLS spectra turns out to be negligible [31] [44].

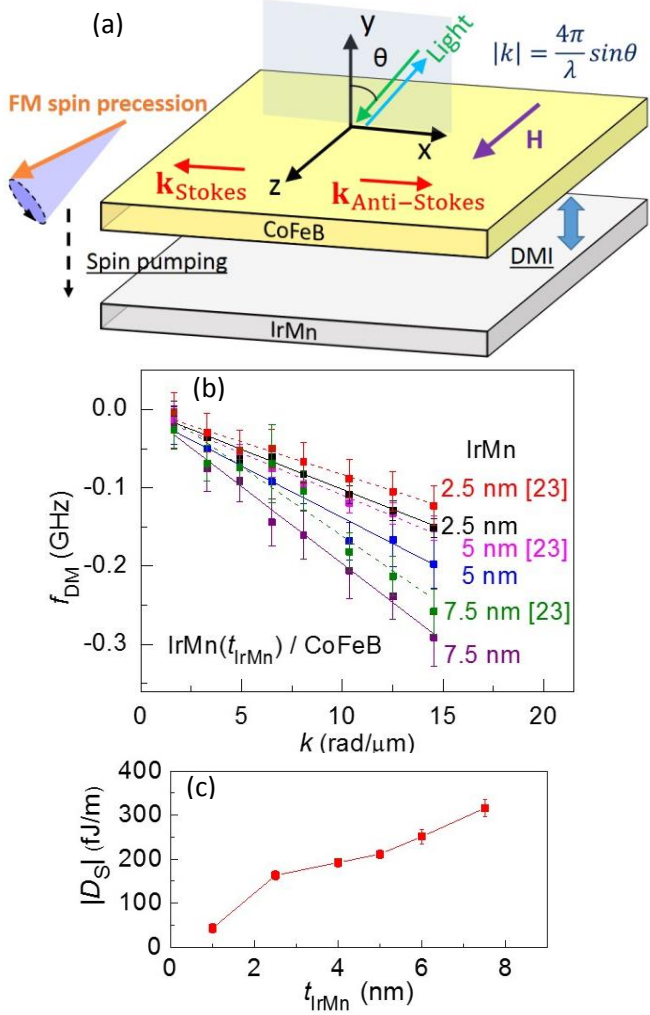

Fig. 1. (a) Schematics of BLS experiment. (b) The linear dependence of DMI induced frequency shift on $k$ for several IrMn(t)/CoFeB(2) samples. The dashed lines show our previous results in Ref. [23] for comparison. (c) The interfacial DMI strength as a function of IrMn thickness in IrMn/CoFeB.

Figure 1b displays the DMI measurement results for certain IrMn/CoFeB samples with different $t_{IrMn}$, where the slope of such linear dependence is used to determine the DMI strength [24, 39-43]. Compared with Ref [23], the DMI strength enhances without the post-annealing procedure. This is likely due to the suppression of the field cooling induced atomic diffusion at the IrMn/CoFeB interface, where stronger DMI benefits from better interface quality [45]. We note that the annealing's impact on DMI may also be of technology relevance to optimize DMI and EB simultaneously, since EB is often controlled by varying the field cooling conditions in AFM/FM systems [6]. Moreover, the DMI strength $|D_S|$ keeps increasing with $t_{IrMn}$ from 2.5 to 7.5 nm, as summarized in Fig. 1c.

We determined the values of $g_{eff}^{\uparrow\downarrow}$ and $\alpha_{sp}$ through the full width half maximum (FWHM) of the BLS spectra, similar to ferromagnetic resonance (FMR) experiments [31, 46]. Figure 2a presents some examples of BLS spectra obtained in the IrMn(5)/CoFeB(2) thin film under different external magnetic, where FWHM increases with larger $H$. Figure 2b plots the BLS linewidth FWHM as a function of $H$, which can be well fitted with

$$\text{FWHM} = \delta f_H + \delta f_0 = \frac{\alpha \gamma}{\pi} H + \delta f_0 \tag{1}$$

Here, the offset $\delta f_0$ is the extrinsic linewidth and unrelated to $H$, resulting from the sample inhomogeneity and instrument build-in linewidth of the interferometer. The slope of the linear dependence is used to estimate the Gilbert damping $\alpha$ of the CoFeB layer with $\alpha = \alpha_{sp} + \alpha_0$. The $\alpha_0$ denotes the intrinsic Gilbert damping of CoFeB layer, which describes the energy flow rate from spin to electronic orbital and phonon degrees of freedom through electron scatterings without the IrMn layer [47]. The value of $\alpha_0$ is estimated by measuring a sample MgO/CoFeB/MgO in Fig. 3a, where $\alpha_{sp} = 0$. The $\alpha_{sp}$ represents the extrinsic Gilbert damping due to the non-local spin relaxation from spin pumping effect at CoFeB/IrMn interfaces. With the value of $\alpha_{sp}$, we further determined the effective spin mixing conductance using $g_{eff}^{\uparrow\downarrow} = \frac{4\pi M_S t_{FM}}{\gamma \hbar} \alpha_{sp}$ at the IrMn/CoFeB interfaces.

nominal layer thicknesses in nanometers. We used thermally oxidized Si substrates with around 100 nm $SiO_2$ on surface, because the light signal is optimized for all incident angle used in BLS [36]. Different from Ref. [23], no annealing treatment was applied after the sputtering procedure. The IrMn layer is poly-crystalline and with a strong (111) peak in the X-ray diffraction results (Fig. S1 [37]). SOT measurements [15, 16] and neutron diffraction studies [38] on similar samples suggest a non-collinear AFM spin alignment in the IrMn layer.

BLS measurements were performed to determine both the DMI and the effective spin mixing conductance $g_{eff}^{\uparrow\downarrow}$ at IrMn/CoFeB interfaces. We used the backscattering geometry shown in Fig. 1a to investigate the thermal magnon spectra of CoFeB. An in-plane magnetic field **H** was applied along the z axis. A laser beam with s-linear polarization was incident on the sample, and the p-polarized component of the backscattered light was

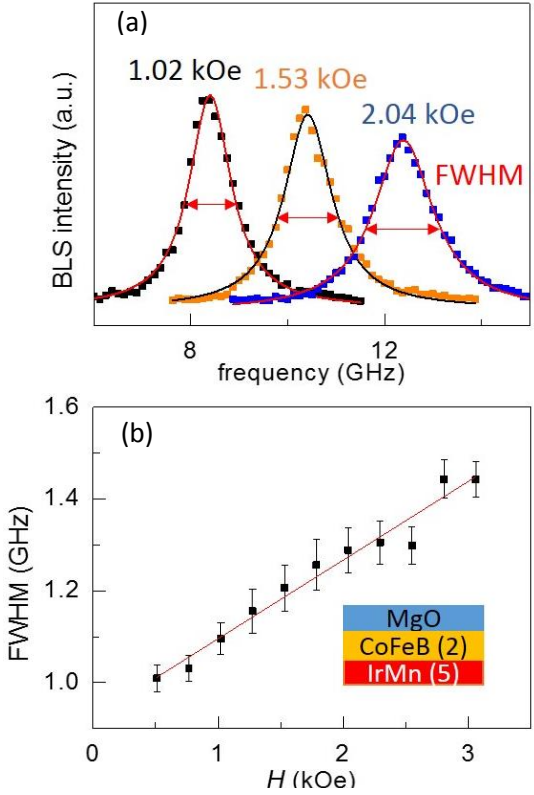

Fig. 2. (a) BLS spectra for DE spin waves recorded at a fixed incident angle with $\theta = 45°$ under different external magnetic fields **H** in the IrMn(5)/CoFeB(2)/MgO sample. The solid lines represent Lorentizian fittings. (b) The linear dependence of FWHM on $H$ in IrMn(5)/CoFeB(2). The solid line refers to the least square fitting.

To understand the intriguing increase of DMI with IrMn thickness $t_{IrMn}$ up to 7.5 nm, we characterized $g_{eff}^{\uparrow\downarrow}$ on the samples with different $t_{IrMn}$. Figure 3a shows the measured $\delta f_H$ as a function of $H$ on different IrMn($t_{IrMn}$)/CoFeB(2) thin films, and the slopes of the linear fittings are summarized in Fig. 3b. Different slopes mainly result from the modification of Gilbert damping $\alpha$, since other physical parameters remain almost unchanged in the thin films [23]. The bump in Fig. 3b near $t_{IrMn} = 1$ nm is likely due to the additional enhancement of $\alpha$ induced by the fluctuations of the magnetic order in the IrMn layer near its transition temperature, which has been demonstrated by previous FMR studies on IrMn/Cu/NiFe thin films with comparable IrMn thicknesses [9]. More important are the results that $\alpha$ keeps increasing with larger $t_{IrMn}$ at $2.5\ nm \leq t_{IrMn} \leq 7.5\ nm$, where DMI at IrMn/CoFeB interface exhibits puzzling difference from that at HM/FM interface [23].

Our key finding is that both $|D_S|$ and $g_{eff}^{\uparrow\downarrow}$ simultaneously increase with IrMn layer from 2.5 nm to 7.5 nm, as plotted in Fig. 4a. We use such correlation to elucidate the underlying physics of DMI at IrMn/CoFeB interface, with the help of the understandings on magnetic damping and spin pumping effect.

In the following discussion, we resolve such AFM/FM interfacial coupling into the impact on FM constituent and the unique role played by the AFM constituent.

The simultaneous increase of $|D_S|$ and $g_{eff}^{\uparrow\downarrow}$ likely originates from the facilitated spin-flip transitions between $3d$ states in the FM CoFeB layer. We elaborate such interpretation by connecting several studies in different topics. On one hand, spin-flip excitations contribute significantly to Gilbert damping in ferromagnetic metals and alloys at room temperature, as a result of the interband electron transitions [47-49]. It has also been demonstrated that the spin pumping enhanced damping $\alpha_{sp}$ depends crucially on the spin flipping at HM/FM interfaces [50]. Therefore, the measured increases of $g_{eff}^{\uparrow\downarrow}$ and $\alpha_{sp}$ reflect that spin-flip transitions are facilitated in CoFeB layer when adjacent to thicker IrMn layer. On the other hand, such processes likely promote larger interfacial DMI. That's because DMI is driven by the spin-flip transitions between $3d$ states (in FM) that involve intermediate states (from the adjacent layer) with strong SOC strength, as demonstrated at HM/FM interfaces [51]. The situation may be similar for the DMI at IrMn/CoFeB interface as illustrated in Fig. 4b, which results in the observed correlation between DMI and $g_{eff}^{\uparrow\downarrow}$ with larger $t_{\text{IrMn}}$. The role of spin-flip transitions on the correlation between $|D_S|$ and $g_{\text{eff}}^{\uparrow\downarrow}$ is consistent with our previous study in the HM/FM systems by varying the HM type [44].

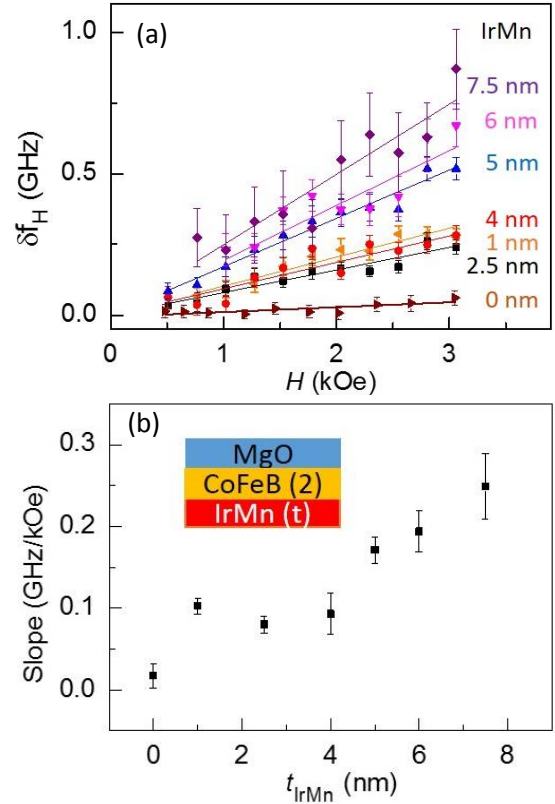

Fig. 3. (a) The linear dependence of FWHM on $H$ in IrMn/CoFeB(2) thin films with different IrMn thicknesses. The "0 nm IrMn" denotes the results obtained from the control sample MgO/CoFeB/MgO. (b) The slopes of such linear correlations change with different IrMn thicknesses.

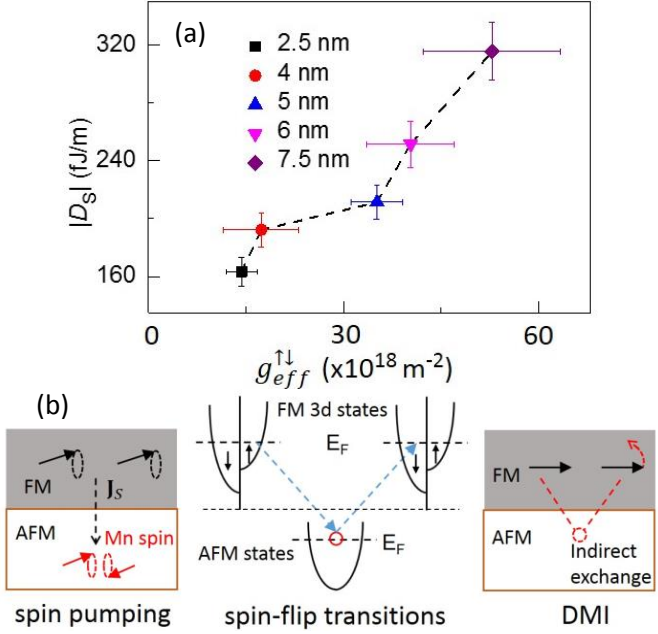

Fig. 4. (a) A correlation between $|D_S|$ and $g_{eff}^{\uparrow\downarrow}$ in IrMn/CoFeB thin films with different IrMn thicknesses. The dashed line serves as visual guide (b) Illustration of a possible interpretation on the observed correlation. The contributions from Ir in IrMn to spin pumping and DMI are not highlighted in the illustration.

Next, we discuss the unique role played by the IrMn layer, leading to the difference of DMI at AFM/FM interfaces from that in HM/FM systems. While the intermediate SOC states are necessary in the spin-flip processes for DMI at both AFM/FM and HM/FM interfaces, these active states of IrMn near Fermi level include not only Ir $5d$ states but also Mn $3d$ ones. The Mn states holding AFM spins may also help facilitate the spin-flip transitions between Co(Fe) $3d$ states through orbital hybridization, and hence contribute to the DMI at IrMn/CoFeB interface as reflected by the opposite DMI signs between Ir/CoFeB and IrMn/CoFeB interfaces [23]. Moreover, we articulate that the surprising increase of DMI with larger $t_{\text{IrMn}}$ is owing to the Mn states' contribution. With thicker IrMn layer, the SOC strength associated with the Mn states is enhanced. Such modification can be inferred from the enlarged AFM anisotropy [6] [52], which subsequently results in a faster dissipation of spin angular momentum to the AFM lattice [12, 53] as manifested by the increases of $g_{eff}^{\uparrow\downarrow}$ and $\alpha_{sp}$. Since DMI scales with the SOC strength of the intermediate states [25], such modification in Mn states by IrMn thickness further enhances the DMI strength beyond the anticipated saturation (i.e., IrMn's spin diffusion length ~ 0.7 nm [53], the value of which is based on an analog to the DMI's dependence on HM thickness in HM/FM systems [24]).

Finally, we show that the above $3d$(CoFe)-$5d$/$3d$ (Ir/Mn)-$3d$ (CoFe) electron hopping procedure contributes to interfacial DMI, only if the IrMn layer is in immediate contact with the CoFeB layer. This is rather different from certain coupling mechanisms between AFM and FM layers, such as that the exchange bias merely disappears in CoO/Cu(t)/Fe thin films with larger than 3.5 nm Cu insertion [54]. One clear evidence is that DMI strength diminishes by inserting 1 nm Cu in between IrMn and CoFeB layers [55], as shown in Fig. S5 [37]. No significant spin relaxation is expected in transversing the 1 nm Cu spacer between the IrMn and CoFeB layers. The drastically reduced DMI results from disrupted hybridization between the $3d$ (CoFe) and $5d$/$3d$ (Ir/Mn) orbitals, where the spatial overlap between those orbitals is crucial.

In conclusion, we characterized the effective spin mixing conductance $g_{\text{eff}}^{\uparrow\downarrow}$ at IrMn/CoFeB interface, and observed a correlation between $g_{\text{eff}}^{\uparrow\downarrow}$ and interfaical DMI with larger IrMn thickness. Such correlation sheds light on that DMI can be enhanced through the enlarged SOC strength in Mn states of IrMn near Fermi level and their facilitation on spin-flip transitions in the CoFeB layer. This finding may provide a new route to strengthen DMI for engineering chiral spin textures such as magnetic skyrmions. We also anticipate that the knowledge of such correlation at AFM/FM interfaces will help guide future AFM spintronic designs, where both DMI and magnetic damping play important roles, as is the case for spin-orbit-torque driven magnetization switching or auto-oscillation in AFM/FM heterostructures.

* Email address: xma518@utexas.edu

**ACKNOWLEDGEMENTS**

We acknowledge the anonymous reviewers for their insightful comments. We also acknowledge Stephen S. Sasaki and Sarah H. Tolbert for their help in field cooling certain samples. The collaboration between UT-Austin and UCLA are supported by SHINES, an Energy Frontier Research Center funded by the U.S. Department of Energy (DoE), Office of Science, Basic Energy Science (BES) under award # DE-SC0012670. Guoqiang Yu acknowledges the financial support from the National Natural Science Foundation of China (NSFC)-Science Foundation Ireland (SFI) Partnership Programme [Grant No. 5171101593]